\def\dev{\partial}
\def\G{\Gamma}
\def\m{\mu}
\def\n{\nu}

\def\l{\lambda}
\def\T{\Theta}

\def\pmb#1{\setbox0=\hbox{#1}%
  \kern-.025em\copy0\kern-\wd0
  \kern.05em\copy0\kern-\wd0
  \kern-.025em\raise.0433em\box0}

\def\bfxi{\pmb{$\xi$}}
\documentstyle[aps,preprint]{revtex}
\topmargin = - 0.9 cm
\begin{document}
\pagestyle{empty}
\preprint{IFUP-TH-18/93}
\vskip 2.5truecm
\title{
\bf  STATIONARY SOLUTIONS AND CLOSED TIME-LIKE CURVES
 IN 2+1 DIMENSIONAL GRAVITY\footnote{Work partially
supported by M.U.R.S.T.}}

\author{P. Menotti}
\address{
Dipartimento di Fisica della Universit\`a, Pisa 56100,
 Italy and\\
INFN, Sezione di Pisa
}
\author{D. Seminara}
\address{
Scuola Normale Superiore, Pisa 56100, Italy and\\
INFN, Sezione di Pisa
}
\maketitle
\begin{center}
May  1993
\end{center}
\newpage
\pagestyle{plain}

\begin{abstract}
We give the general solution of the stationary problem of 2+1
dimensional gravity in presence of extended sources, also endowed
with angular momentum. We solve explicitly the compact support
property of the energy momentum tensor and we apply the results to
the study of closed time-like curves. In the case of rotational
symmetry we prove that the weak energy condition combined with the
absence of  closed time-like curves at space infinity prevents the
existence of closed time-like curves everywhere in an open universe
(conical space at infinity).
\end{abstract}
\section{Introduction}

A great deal of interest has been devoted to 2+1 dimensional gravity
\cite{flat},
more recently in connection with the problem of closed time-like curves
(CTC)
\cite{Gott,JDtH,GuthI,GuthII,Cutler,Kabat,
Ori,t Hooft,Tipler,Hawking,DJ,Grant,MS3}. The
 aspect of exact solutions in 2+1 dimensional gravity has been
addressed in detail, both in the case of point like sources \cite{flat}
 and in the
case of extended sources \cite{MS1,MS2}.
Some inroads have also been done in the
realm of time dependent solutions \cite{hooft,MS1,MS2}. The radial gauge
approach \cite{MT,MS1,MS2}
has proven particularly fruitful in obtaining quadrature formulae
for extended and non stationary sources.
On the other hand most of the activity has been devoted to point like
sources where the energy momentum tensor acquires a particularly
simplified form. This has been done both
for sources without and with spin; however point like sources with
spin have been generally considered unphysical as they generate in
their proximity metrics that support CTC. From this viewpoint considering
extended sources is of great interest. In a previous paper \cite{MS2}
we gave in
terms of quadratures the
general resolvent formulae for the time dependent solutions, and we
solved completely the support problem in the case of rotational symmetry.
Obviously finding solutions of Einstein's equations whose energy momentum
tensor has some
prescribed support
and symmetry properties is not enough as the energy momentum tensor
should satisfy some
energy condition \cite{H.E.}. This is not a trivial problem
as it involves inequalities among the eigenvalues
of the energy momentum tensor in the metric generated,
through Einstein's equations, by
the energy momentum tensor itself.
As we shall see the radial gauge, due to its physical nature, provides a
powerful device for extracting useful information from the energy
condition and in constructing sources that satisfy such energy conditions.
In this paper we shall produce resolvent formulae, in term of
quadratures for the general stationary problem with extended sources and
give a complete treatment of the CTC that may appear in the case of
rotational symmetry. One of the main results of the paper will be that
the imposition of ~~i)
the weak energy condition (WEC) and~~ ii) the absence of
CTC at space infinity, prevents the occurrence of CTC anywhere in an
open (conical) universe. To understand how far both conditions are
also necessary to avoid CTC we provide solved examples of
regular sources with non vanishing total angular momentum, which
violate
WEC, have no CTC at infinity but produce CTC for some finite radius.
In addition the
assumption that no CTC appears at space infinity is a necessary one, as we
shall produce solved examples of regular sources that satisfy the WEC
but produce CTC at infinity.

The paper is organized as follows: in sect.II we introduce the reduced
radial gauge, which will be the main tool in the sequel of the paper,
and write down Einstein's equations in such a gauge for the most
general stationary metric. In sect.III we give the resolvent formulae
in terms of quadratures, that express the metric in terms of the
energy momentum tensor, giving the explicit condition for the
compactness of the support of the energy momentum tensor in the
general case. We discuss also the simplifying features that occur in
the case of rotational symmetry. In sect.IV we address the problem of
CTC for stationary solutions with rotational symmetry and prove the
main result that the WEC combined with the absence of CTC at space
infinity prevents the occurrence of CTC everywhere. In sect.V we
summarize the main conclusion and outline possible developments.
In appendix A we
derive the resolvent formulae for the reduced radial gauge, in
appendix B we give the general regularity conditions for the energy
momentum tensor at the origin and in appendix C
we give the derivation of the compact support condition for
the energy momentum tensor in the general case.

\section{Reduced radial gauge}
In stationary  problems the reduced radial gauge \cite{MS1} is defined by
\begin{equation}
\sum_i\xi^i\G^a_{bi}=0,
\end{equation}
\begin{equation}
\sum_i\xi^i e^a_i=\sum_i\delta^a_i\xi^i,
\end{equation}
where the sums run over the space indices. (In the following the indices $i
,\ j,\ k,\ l,\ m$ denote space indices).

It was shown in ref. \cite{MS1} sect.4 that such a gauge is attainable
and that the reference frame that realizes it  is the (generalized)
Fermi-Walker coordinate system for an observer  that moves along an
integral line of the time-like Killing vector field.

In any space-time dimension there are resolvent formulae that express the
vierbeins and the connection  in terms of the Riemann and torsion
tensor in this gauge. They are (see appendix A)

\begin{equation}
\Gamma^a_{bi}({\bfxi})=\xi^j \int^1_0 R^a_{bji}(\l{\bfxi})\l d\l,
\end{equation}
\begin{equation}
\Gamma^a_{b0}({\bfxi})=\G^a_{b0}({\bf 0})+\xi^i\int^1_0 R^a_{bi0}(\l {\bfxi})
d\l,
\end{equation}
\begin{equation}
\label{eai}
e^a_i=\delta^a_i+\xi^j\xi^l\int^1_0 R^a_{jli}(\l{\bfxi})\l(1-\l) d\l+
\xi^j\int^1_0 S^a_{ji}(\l{\bfxi})\l d\l,
\end{equation}
\begin{equation}
\label{ea0}
e^a_0=\delta^a_0+\xi^i\G^a_{i0}({\bf 0})+\xi^i\xi^j\int^1_0 R^a_{ij0}
(\l{\bfxi})(1-\l) d\l+\xi^j\int^1_0 S^a_{j0}(\l{\bfxi}) d\l.
\end{equation}
In the following we shall consider theories with vanishing torsion.
The peculiarity of 2+1 dimensions is the substantial identification of the
Riemann with the Ricci tensor and through Einstein's equation, with the
energy momentum tensor. Explicitly
\begin{equation}
\label{einstein}
\varepsilon_{abc} R^{ab}= -{2\kappa} T_c,
\end{equation}
where $\kappa=8\pi G$, and thus
\begin{equation}
R^{ab}=-{\kappa}\varepsilon^{abc} T_c=-{\kappa\over 2}
\varepsilon^{abc}~\varepsilon_{\rho\m\n}\tau^{~\rho}_c dx^\m\wedge dx^\n,
\end{equation}
where $T_c$ is the energy momentum two form. Using such a relation one can
express through  a simple quadrature, the connections and  the
vierbeins in terms of the energy momentum tensor, which is the source of
the gravitational  field and thus one solves Einstein's equation.
On the other hand the energy momentum tensor
  is  subject to the covariant conservation
law and  symmetry condition. Thus our problem is to construct
the general form of
conserved symmetric sources in the reduced radial gauge,  which in addition
should satisfy other physical requirement given by the support of the
sources and the restrictions due to  the energy condition \cite{H.E.}.

The conservation and symmetry equations for the energy momentum tensor are
\begin{equation}
{\cal D} T^a=0,
\end{equation}
\begin{equation}
\label{symmetry}
\varepsilon_{abc} T^b\wedge e^c =0.
\end{equation}
To solve these equations  we use the technique  developed in \cite{MS2}.
It is useful  to introduce  the  cotangent vectors
$\displaystyle{T_{\mu}=\frac{\partial \xi^0}{\partial \xi^
\mu}}$,
$\displaystyle{P_{\mu}=\frac{\partial \rho}{\partial \xi^
\mu}}$ and
$\displaystyle{\Theta_{\mu}=\rho\frac{\partial \theta}{\partial \xi^
\mu}}$ where $\rho$ and  $\theta$ are the polar variables in the
$(\xi^1,\xi^2)$ plane.
The conservation equation is solved by the general ansatz \cite{MS1}
\begin{eqnarray}
\label{e.m.t.}
\tau^\rho_c(\bfxi)&&=\frac{1}{\kappa} \biggl [ P^\mu \partial_\mu A^\rho_c
(\bfxi)
-\frac{1}{\rho} A^\rho_c (\bfxi)-\frac{1}{\rho} \Theta^\rho \Theta_\mu
A^\mu_c(\bfxi)- P^\rho \biggl (\partial_\mu  A^\mu_c(\bfxi)-\nonumber\\
&&-\frac{1}{2}\varepsilon_{clm}\varepsilon_{\alpha\beta\sigma} P^\alpha
 A^{l\beta}(\bfxi) A^{m\sigma}(\bfxi)\biggr )\biggr ],
\end{eqnarray}
where $A^\rho_c(\bfxi)$ is related  to the connection
$\Gamma^{a}_{b\mu}(\bfxi)$ in
the reduced radial gauge  in $2+1$ dimensions by
\begin{equation}
\label{connection}
\Gamma^{ab}_\mu(\bfxi)=\varepsilon^{abc}\varepsilon_{\mu\rho\nu}P^\rho
A^\nu_c(\bfxi).
\end{equation}
The origin of (\ref{e.m.t.}) is the following: if we express through
Einstein's equations (\ref{einstein}) the energy momentum tensor in
terms of the connection written as in (\ref{connection}), which is the most
general form for a radial connection in 2+1 dimensions, we obtain
(\ref{e.m.t.}).
The most general form of $A^\rho_c(\bfxi)$, taking into account that the
components of $A^\rho_c$
 along $P^\rho$ are irrelevant to determine the geometry, is
\begin{equation}
\label{A}
A^\rho_c(\bfxi)= T_c \left [ \Theta^\rho\beta_1+T^\rho\frac{(\beta_2-1)}{\rho}
\right ]+\Theta_c \left [\Theta^\rho\alpha_1 +T^\rho\frac{\alpha_2}{\rho}
\right ]+P_c \left [\Theta^\rho \gamma_1+T^\rho\frac{\gamma_2}{\rho}
\right ].
\end{equation}
We use for the coefficient of $T_c T^\rho$ the form  $\beta_2-1$,
as it simplifies the writing of subsequent formulae.
Substituting (\ref{A}) into (\ref{e.m.t.}) we obtain
\begin{eqnarray}
\label{taumisto}
&&\tau^{\rho}_c=-\frac{1}{\kappa}\biggl \{ T_c \biggl ( T^\rho
\frac{\beta^\prime_2}{\rho}+\T^\rho \beta^\prime_1\biggr )+
\T_c
\biggl ( T^\rho
\frac{\alpha^\prime_2}{\rho}+\T^\rho\alpha^\prime_1
\biggr )+
 P_c\biggl (T^\rho \frac{\gamma^\prime_2}{\rho}+
\T^\rho
\gamma^\prime_1\biggr ) +\\
&&\frac{1}{\rho}P^\rho
\biggl [ T_c\biggl (\alpha_1\gamma_2-\alpha_2\gamma_1-
\frac{\dev\beta_1}{\dev \theta}\biggr )+
\T_c \biggl (\beta_1\gamma_2-\beta_2\gamma_1-
\frac{\dev\alpha_1}{\dev \theta}\biggr )
+P_c \biggl (\alpha_1\beta_2
-\alpha_2\beta_1-
\frac{\dev\gamma_1}{\dev \theta}\biggr )\biggr ]\biggr \}.\nonumber
\end{eqnarray}
Using eq. (\ref{einstein}) and (\ref{taumisto}) into
(\ref{eai}) and (\ref{ea0}) we obtain
\begin{equation}
\label{drei1}
e^a_0(\bfxi)=-T^a A_1 -\Theta^a  B_1,
\end{equation}
\begin{equation}
\label{drei2}
e^a_i(\bfxi)=-\frac{1}{\rho}\Theta^a\Theta_i B_2-
\frac{1}{\rho} T^a \Theta_i A_2- P^a P_i,
\end{equation}
where $A_i$ and $B_i$ are defined by
\begin{eqnarray}
&&A_1(\bfxi)=\rho \int^1_0 \alpha_1(\lambda \bfxi) d\lambda-1\ \ {\  ,\  }
\ \ \ B_1(\bfxi)=\rho \int^1_0 \beta_1(\lambda \bfxi) d\lambda \nonumber,   \\
&&A_2(\bfxi)=\rho \int^1_0 \alpha_2(\lambda \bfxi) d\lambda\phantom{-1}
\ \ {\rm and}
\ \ \ B_2(\bfxi)=\rho \int^1_0 \beta_2(\lambda \bfxi) d\lambda.
\end{eqnarray}
Eqs. (\ref{drei1}) and (\ref{drei2}) can be summarized into
\begin{equation}
\label{dreibein}
e^a_\mu(\bfxi)= -T^a (T_\mu A_1+\frac{1}{\rho} \Theta_\mu A_2)-
\Theta^a (T_\mu B_1+\frac{1}{\rho} \Theta_\mu B_2)-P^a P_\mu
\end{equation}
from which the metric is given by
\begin{equation}
ds^2=
(A_1^2-B_1^2) dt^2
+ 2 (A_1 A_2-B_1 B_2)dtd\theta + (A_2^2-B^2_2) d\theta^2-d\rho^2.
\end{equation}
Contracting the energy momentum tensor (\ref{taumisto}) with the
dreibein  (\ref{dreibein})
we obtain the same quantity in the dreibein base
\begin{eqnarray}
\label{tab}
&&\tau^{c\mu}(\bfxi) e^a_\mu(\bfxi)={\cal T}^{ca}=
-\frac{1}{\kappa \rho}\biggl \{
T^a \biggl [ T^c \biggl (A_2 \beta^\prime_1-A_1 \beta^\prime_2 \biggr )+
 \Theta^c \biggl ( A_2 \alpha_1^\prime-A_1\alpha_2^\prime\biggr )+
\nonumber \\
&&P^c\biggl (A_2 \gamma_1^\prime-A_1\gamma_2^\prime\biggr )\biggr ] +
\Theta^a \biggl [ T^c \biggl (B_2 \beta^\prime_1-B_1 \beta^\prime_2
\biggr )+
\Theta^c \biggl ( B_2 \alpha_1^\prime-B_1\alpha_2^\prime\biggr )+
\nonumber \\
&&P^c\biggl (B_2 \gamma_1^\prime-B_1\gamma_2^\prime\biggr )\biggr ]+
P^a \biggl [ T^c \biggl (\alpha_1 \gamma_2-\alpha_2 \gamma_1-\frac{\partial
\beta_1}{\partial \theta} \biggr )+
\Theta^c \biggl (\beta_1 \gamma_2-\beta_2 \gamma_1-
\nonumber \\
&&\frac{\partial
\alpha_1}{\partial \theta}  \biggr )+
P^c\biggl (\alpha_1 \beta_2-\alpha_2 \beta_1-\frac{\partial
\gamma_1}{\partial \theta} \biggr )\biggr ]\biggr \}.
\end{eqnarray}
We shall consider solutions of Einstein' s equations free of physical
singularities, which implies that also the energy momentum tensor is
regular.
The regularity of $\Gamma^{ab}_\mu(\bfxi)$ and of $\tau_{c}^\rho(
\bfxi)$ in a neighborhood of the origin impose conditions on the behavior
of $\alpha_i,\ \beta_i,\ \gamma_i$  at the origin. We start from
$\Gamma^{ab}_\mu(\bfxi)$. Its continuous behavior at the origin imposes
the following behavior upon  the basic functions $\alpha_i$, $\beta_i$,
$\gamma_i$
\begin{equation}
\label{reg}
\begin{array}{ccc}
&\alpha_1\to \Gamma^{01}_0({\bf 0}) \cos\theta+\Gamma^{02}_0 ({\bf 0})
\sin\theta &\ \ \ , \  \alpha_2\to o(\rho),\\
&\beta_1\to \Gamma^{12}_0({\bf 0})\phantom{\cos\theta+c_2 si} &
\ \ \ ,\ \ \ \   \phantom{aa}\beta_2\to 1+o(\rho),\\
&\gamma_1\to -\Gamma^{02}_0({\bf 0}) \cos\theta+\Gamma^{01}_0({\bf 0})
 \sin\theta  & \ \ ,\ \ \gamma_2\to o(\rho).
\end{array}
\end{equation}
Eqs. (\ref{reg}) do not yet imply the regularity at the origin of the energy
momentum tensor, which, containing derivatives of $\Gamma^{ab}_\mu$,
imposes stronger restrictions. Due to the
singular nature of polar coordinates at the origin, it is more proper
to discuss the regularity at the origin of the energy momentum tensor
in carthesian coordinates and they are given in appendix B.

Coming back to the energy momentum tensor ${\cal T}^{ab}$ the symmetry
constraint (\ref{symmetry}) gives three differential relations
\begin{mathletters}
\begin{eqnarray}
\label{22a}
&&A_1\alpha_2^\prime -A_2 \alpha_1^\prime
+B_2\beta_1^\prime
 -B_1 \beta_2^\prime=0\\
\label{22b}
&&\alpha_2 \gamma_1-\alpha_1 \gamma_2+A_2\gamma_1^\prime -A_1
\gamma_2^\prime +\frac{\partial\beta_1}{\partial\theta}=0\\
\label{22c}
&&\beta_2 \gamma_1-\beta_1 \gamma_2+B_2\gamma_1^\prime -B_1
\gamma_2^\prime +\frac{\partial\alpha_1}{\partial\theta}=0.
\end{eqnarray}
\end{mathletters}
These relations can be integrated with respect to $\rho$ and taking into
account the regularity conditions of $\alpha_i$, $\beta_i$ and $\gamma_i$
at the origin we reach
\begin{mathletters}
\label{symmetryeq}
\begin{eqnarray}
\label{symmetry1}
&&A_1\alpha_2-A_2 \alpha_1+B_2
\beta_1-B_1\beta_2=0\\
\label{symmetry2}
&&A_2\gamma_1 -A_1
\gamma_2 +\frac{\partial B_1}{\partial\theta}=0\\
\label{symmetry3}
&&B_2\gamma_1-B_1
\gamma_2 +\frac{\partial A_1}{\partial\theta}=0.
\end{eqnarray}
\end{mathletters}
In general, in absence of rotational symmetry, caustics may develop in
the sense that geodesics emerging from the origin with different
$\theta$ can intersect at some point for large enough $\rho$. This renders
the map of $\rho,\theta$ into the physical points of space not one to
one, but the geometry can be still regular in the sense that a proper
change of coordinates removes the singularity. For an example of how
this non single valuedness can show up and how it can be removed by changing
coordinates, we
refer to the appendix  of ref. \cite{MS1}. Such a
problem does not arise in the case of rotational symmetry, that  we
discuss in sec. III B  and apply in sect.IV.

\bigskip
\section{Support Properties of the Energy Momentum Tensor}
\subsection{Generic case}
First we give the solution of eqs. (\ref{symmetryeq}) in the generic
case i.e.  when $\alpha_1$ and $\beta_1$ depend on $\theta$ (absence
of rotational symmetry). In the system (\ref{symmetryeq}) we have 6
 unknown functions $\alpha_i,\ \beta_i,\ \gamma_i$ ($i=1,2$)
and 3 equations. We shall choose as independent functions
$\alpha_1,\ \beta_1,\ \gamma_1$. This choice
simplifies the discussion of the support property of the energy
momentum tensor. The method of solution follows the one described in
ref. \cite{MS2} for the complete radial gauge. Multiplying
(\ref{symmetry2}) by $B_1$ and (\ref{symmetry3}) by $A_1$ and subtracting
we obtain
\begin{equation}
\label{N}
A_2 B_1 -A_1 B_2=\frac{1}{2\gamma_1}\frac{\partial}{\partial\theta}
(A_1^2-B_1^2)\equiv N(\rho , \theta)
\end{equation}
from which
\begin{equation}
B_2=\frac{A_2}{A_1} B_1-\frac{N}{A_1}
\end{equation}
and
\begin{equation}
\label{pippo}
\beta_2=\frac{\partial }{\partial\rho}\left (\frac{A_2}{A_1}
\right ) B_1+\frac{A_2}{A_1}\beta_1
-\frac{\partial }{\partial\rho}\left (\frac{N}{A_1}
\right).
\end{equation}
Substituting into eq. (\ref{symmetry1}) we obtain
\begin{equation}
\label{Eq27}
\frac{\partial }{\partial\rho}\left (\frac{A_2}{A_1}
\right )=\frac{B^2_1}{B_1^2-A^2_1}
\frac{\partial }{\partial\rho}\left (\frac{N}{A_1 B_1}
\right)
\end{equation}
from which
\begin{equation}
\label{A2}
A_2=\frac{N B_1}{B^2_1-A^2_1}+2 A_1 I
\end{equation}
where
\begin{equation}
\label{I}
I=\int^\rho_0 d\rho^\prime \frac{N (A_1\beta_1-B_1 \alpha_1)}{(B_1^2-
A^2_1)^2}
\end{equation}
and
\begin{equation}
\label{a2}
\alpha_2=\frac{B^2_1}{B_1^2-A^2_1}
\frac{\partial }{\partial\rho}\left (\frac{N}{ B_1}\right )+
2\alpha_1 I.
\end{equation}
Substituting $\alpha_2$ and $A_2$ thus obtained into (\ref{pippo})
we have
\begin{equation}
\label{b2}
\beta_2=\frac{A^2_1}{B_1^2-A^2_1}
\frac{\partial }{\partial\rho}\left (\frac{N}{ A_1}\right )+
2\beta_1 I,
\end{equation}
and finally using e.g. (\ref{symmetry2}) we have
\begin{equation}
\label{g2}
\gamma_2=\frac{B^2_1}{B_1^2-A^2_1}
\frac{\partial }{\partial\theta}\left (\frac{A_1}{ B_1}\right )+
2\gamma_1 I.
\end{equation}
Formulae (\ref{a2}), (\ref{b2}) and (\ref{g2}) give the solution of
the problem  in terms of the single quadrature $I$.\\
We must now discuss the support property of the source i.e.  the
necessary and sufficient conditions that $\alpha_1,\ \beta_1,\ \gamma_1$
have to satisfy in order that the energy momentum tensor vanishes outside a
certain boundary $\rho_0(\theta)$. From eq. (\ref{taumisto}) the vanishing
of $\tau^{\rho}_a$ imposes
\begin{equation}
\label{ridcompattezza}
\frac{\partial\alpha_i}{\partial \rho}=
\frac{\partial\beta_i}{\partial \rho}=
\frac{\partial\gamma_i}{\partial \rho}=0 \ \ \ {\rm for} \ \ \ \rho>\rho_0(
\theta)
\end{equation}
and
\begin{mathletters}
\label{CC}
\begin{eqnarray}
\label{ridcomp3}
&& \alpha_1\beta_2 -\alpha_2
\beta_1 -\frac{\partial \gamma_1}{\partial\theta}=0\\
\label{ridcomp1}
&&\alpha_2\gamma_1 -\alpha_1
\gamma_2 +\frac{\partial \beta_1}{\partial\theta}=0\\
\label{ridcomp2}
&&\beta_2\gamma_1-\beta_1
\gamma_2 +\frac{\partial \alpha_1}{\partial\theta}=0
\end{eqnarray}
\end{mathletters}
for $\rho>\rho_0(\theta)$. If $\alpha_1,\ \beta_1,\ \gamma_1$
satisfy eqs. (\ref{22b}), (\ref{22c}) and
condition (\ref{ridcompattezza}), also
(\ref{ridcomp1}) and (\ref{ridcomp2}) are  satisfied. Thus we must
simply impose  (\ref{ridcompattezza}) and (\ref{ridcomp3}). In appendix C
we prove that necessary and sufficient condition for this to happen is that\\
\begin{equation}
\label{support1}
\alpha_1 B_1- A_1 \beta_1={\rm constant}\ \ \ {\rm  for}
\ \ \rho>\rho_0(\theta)
\end{equation}
\noindent
and
\begin{equation}
\label{support2}
\alpha^2_1 - \beta^2_1 + \gamma^2_1={\rm constant}\ \ \ {\rm  for}
\ \ \rho>\rho_0(\theta)
\end{equation}
where the two constants do not depend both on $\theta$ and on $\rho$.
Thus given  $\alpha_1$ and $\beta_1$ that satisfy (\ref{ridcompattezza})
and (\ref{support1})
one can easily construct a $\gamma_1$ that satisfy (\ref{support2}) and
thus also
$\gamma^\prime_1=0$ for $\rho>\rho_0(\theta)$. There is no problem in
choosing $\alpha_1$ and $\beta_1$ to satisfy (\ref{support1}).
Then the explicit solution is given by
eqs.(\ref{a2}),(\ref{b2}),
(\ref{g2}).
An alternative to the choice $\alpha_1,
\ \beta_1\ {\rm and}\ \gamma_1$ for the
fundamental functions, is  $\alpha_1,\ \beta_1\ {\rm and}\ N=A_2 B_1-A_1
B_2={\rm det}(e)=\rho~{\rm det}(e^a_\mu)$, where ${\rm det}(e)$ and
${\rm det}(e^a_\mu)$  are the determinants of the dreibeins in polar
and carthesian coordinates. Such a choice avoids the necessity of
dividing by $\gamma_1$ in (\ref{N}). Thus  (\ref{a2}),(\ref{b2}) and
(\ref{g2}) hold
also for $\gamma_1=0$ as it happens in  the case
of rotational symmetry. We notice
furthermore that the energy momentum tensor can be written in algebraic
form in terms of  $A_1$, $B_1$ and $N$ and their first and second derivatives.

\subsection{Case of rotational symmetry}
In this case as the function $\alpha_i$, $\beta_i$ and $\gamma_i$ (see
appendix B of \cite{MS2}) do not depend on $\theta$, from the last two
symmetry equations
\begin{eqnarray}
&&A_2\gamma_1 -A_1
\gamma_2 =0\nonumber\\
&&B_2\gamma_1-B_1
\gamma_2 =0
 \end{eqnarray}
we obtain $\gamma_1=\gamma_2=0$, under the assumption that
the determinant of the dreibein
$\displaystyle{{\rm det}(e)=\frac{1}{\rho}( A_2 B_1- A_1 B_2)}$ never vanishes.
Then the energy momentum tensor ${\cal T}^{ca}$ becomes
\begin{eqnarray}
\label{tabrot}
{\cal T}^{ca}=-\frac{1}{\kappa \rho}\biggl \{
&&T^a \biggl [ T^c \biggl (A_2 \beta^\prime_1-A_1 \beta^\prime_2 \biggr )+
 \Theta^c \biggl ( A_2 \alpha_1^\prime-A_1\alpha_2^\prime\biggr )\biggr
]+
\nonumber \\
&&\Theta^a \biggl [ T^c \biggl (B_2 \beta^\prime_1-B_1 \beta^\prime_2
\biggr )+
\Theta^c \biggl ( B_2 \alpha_1^\prime-B_1\alpha_2^\prime\biggr )
\biggr ]+\nonumber \\
&&P^a P^c\biggl (\alpha_1 \beta_2-\alpha_2 \beta_1 \biggr )\biggr \},
\end{eqnarray}
and the metric
\begin{equation}
ds^2= (A_1^2-B_1^2) dt^2 +
2 (A_1 A_2-B_1 B_2)dtd\theta + (A_2^2-B^2_2) d\theta^2-d\rho^2.
\end{equation}
In the case of rotational symmetry the independence on $\theta$ of the
relations given in appendix B gives the following behavior for the
functions $\alpha_i$,$\beta_i$
\begin{equation}
\label{regrot}
\alpha_1=O(\rho),~~~~\alpha_2=o(\rho^2),~~~~\beta_1=c+o(\rho),
{}~~~~\beta_2=1+O(\rho^2).
\end{equation}
It is easily checked that the behavior of eq.(\ref{regrot}) makes the
connection $\Gamma^{ab}_\mu$ regular at the origin.

The vanishing of $\tau^\rho_c$ outside a finite support,
$\rho>\rho_0$, imposes using the
same reasoning as above
\begin{equation}
\label{Comprot1}
\alpha_1^\prime=\beta_1^\prime=\alpha_2^\prime=\beta_2^\prime=0 \ \ \
{\rm for}\ \ \rho>\rho_0
\end{equation}
and
\begin{equation}
\label{Comprot2}
\alpha_1\beta_2-\alpha_2\beta_1=0  \ \ \
{\rm for}\ \ \rho>\rho_0.
\end{equation}
The only surviving symmetry equation is now
\begin{equation}
\label{symmrot}
A_1\alpha_2-A_2 \alpha_1+B_2\beta_1-B_1 \beta_2=0.
\end{equation}
We notice that eq.(\ref{symmrot}) is invariant under the three dimensional
group of transformations
\begin{equation}
\label{group}
\begin{array}{ccc}
\alpha_1\to \alpha_1+\omega_1 \alpha_2, &\ \ \ & \beta_1\to c~ \beta_1+\omega_2
\beta_2\\
\alpha_2\to c~ \alpha_2\phantom{\omega_2c \alpha_2}, &\ \ \ \ \ \ \ &
\beta_2\to \beta_2\phantom{nnc+\omega_2 (\beta_2+1)}.\\
\end{array}
\end{equation}
This  is what is left of the larger  $Sp(2)\times U(1)$
invariance of eq. (\ref{symmrot}) once one imposes that the symmetry
transformation respects the regularity conditions of $\Gamma^{ab}_\mu(\bfxi
)$ at the origin.
We observe that for $\omega_1=\omega_2=\omega$ and $c=1$ the
transformations (\ref{group})
leave ${\cal T}^{ca}$ invariant and as such also the support
equations (\ref{Comprot1}) and (\ref{Comprot2}). It means that
transformation (\ref{group}) with $\omega_1=\omega_2=\omega$ and $c=1$
corresponds to the same physical system described in a different  frame.
In fact the metric differs by a rigid rotation with angular
velocity $\omega$. On the other hand the transformations with $\omega_1
\neq \omega_2$ and/or $c\neq1$ correspond to different physical
situations.

We can exploit the transformation (\ref{group}) to generate all solutions
for energy-momentum tensors with  bounded support. In fact it is
immediately seen that once three among the four functions
$\alpha_i,\ \beta_i$ have zero derivatives for $\rho>\rho_0$, eq.
(\ref{symmrot}) is solved by the remaining one also constant for
$\rho>\rho_0$.  The
problem is to satisfy the bounded support condition (\ref{Comprot2}). We
shall show that all solutions with bounded support, through a rotation with
$\omega_1=\omega_2=\omega$, $c=1$, can be reduced to one of the following
cases, where the bounded support condition is trivially satisfied.

1. $\alpha_2^0$  and/or $\beta_2^0\neq 0$ and $\alpha^0_1=\beta^0_1=0$.

2. $\alpha_2^0=\beta_2^0=0$.

\noindent
In fact if $\alpha^0_2\neq 0$ choosing from (\ref{group})
$\omega=-\alpha^0_1/ \alpha^0_2$ we obtain $\alpha^0_1\rightarrow 0$, but
(\ref{Comprot2}), which is invariant under such a transformation implies
$\beta^0_1\rightarrow 0$. Similarly one reasons for $\beta^0_2\neq 0$.

\noindent
{\bf Case 1.} For ${\alpha_2^0}^2-{\beta_2^0}^2\neq 0$ we have  that
$g_{00}$ and
$g_{0\theta}$  outside the source are constant
while
$g_{\theta\theta}$ behaves like $({\alpha_2^0}^2-{\beta_2^0}^2)\rho^2$. If
$g_{00}>0$ due to eq. (\ref{symmrot}) ${\alpha_2^0}^2-{\beta_2^0}^2<0$
and we have the usual conical
universe with angular deficit
$\delta=2\pi(1-\sqrt{{\beta_2^0}^2-{\alpha_2^0}^2}) $,
related to the source mass by $\delta=8\pi GM$ and angular momentum
$g_{0\theta}/g_{00}=(A^0_1A^0_2-B^0_1B^0_2)/({A^0}^2_1-{B^0}^2_1)=4 G J$. If
$g_{00}={{A_1^0}^2}-{B_1^0}^2<0$ then we must have
${\alpha^0_2}^2-{\beta^0_2}^2>0$ because of eq.(\ref{symmrot}).
 Then by a proper rotation the metric can be reduced to
$$
({\alpha^{0}_2}^2-{\beta^{0}_2}^2)\left (\rho-\rho_0+{\alpha_2^0 A_2^0-
\beta_2^0B_2^0
\over {A_1^{0}}^2-{B_1^{0}}^2}\right )^2\times
$$
\begin{equation}
({{A_1^{0}}^2-{B_1^{0}}^2\over A_1^0 A_2^0-B_1^0 B_2^0}dt - d\theta)^2+
{(A_1^0 A_2^0-B_1^0 B_2^0)^2\over {A_1^{0}}^2-{B_1^{0}}^2} d\theta^2 -d\rho^2
\end{equation}
which is the usual outer metric generated by a closed string with tension
\cite{flat}.
If ${\alpha_2^0}^2={\beta_2^0}^2\neq0$ we have a``linear'' universe in which
$g_{\theta\theta}$ behave linearly for $\rho\rightarrow\infty$(see
\cite{flat}).\\
\noindent
{\bf Case 2.}  $g_{\theta\theta}$ and $g_{0\theta}$ are constant for
$\rho>\rho_0$ and if ${\alpha^0}^2_1-{\beta^0}^2_1\neq 0$,
$g_{00}$ behaves like $\rho^2$ for $\rho\rightarrow\infty$.
 On the other hand
$g_{\theta\theta}\neq 0$ because otherwise ${A^0}_2=\pm {B^0}_2$ which implies
through eq.(\ref{symmrot})
${\alpha^0_1}=\pm{\beta^0_1}$ which contradicts
${\alpha^0}^2_1-{\beta^0}^2_1\neq 0$. ($A_2^0=B^0_2=0$ is excluded
because otherwise for $\rho>\rho_0$ the metric degenerates).
Then by means of a rotation we can set
 $g_{0\theta}=0$ to reach, outside
the source, the metric
\begin{equation}
ds^2=k (c_1+\rho)^2dt^2+g_{\theta\theta}d\theta^2-d\rho^2.
\end {equation}
which for positive $k$ is the
metric generated by a static string with tension \cite{flat}. For
$k<0$ due to the signature we have $g_{\theta\theta}=\displaystyle{-\frac{(A_1
B_2-A_2 B_1)^2}{g_{00}}}>0$ and outside the source the role of the
angular variable is exchanged with that of time.

\noindent
If ${\alpha_1^0}^2={\beta_1^0}^2\not =0$, $g_{00}$ is linear at infinity,
$g_{0\theta}$
is constant and $g_{\theta\theta}$ is zero.
Finally if $\alpha^0_i=\beta^0_i=0$ we have a cylindric universe.

\section{Closed time-like curves and the weak energy condition.}
\noindent
Recently considerable interest has been devoted to the problem of
closed time-like curves (CTC) in 2+1 dimensional gravity. This is of
interest also for the 3+1 dimensions as all solutions in 2+1 can be
considered as solutions of 3+1 dimensional gravity in presence of a
space-like Killing vector field (cosmic strings).
Gott \cite{Gott}
 was able to produce examples of kinematical configurations of point
particles without spin which produce CTC in 2+1 dimensions and
this system came under close
scrutiny \cite{JDtH,GuthI,GuthII,Cutler,Kabat,Ori,t Hooft}.
 Carrol, Fahri and Guth \cite{GuthII} proved that if the universe
containing  spinless point particles is open and has
total time like momentum, no Gott pair can be created during its evolution
thus supporting the idea that in an open time-like universe no CTC can form.
't Hooft \cite{t Hooft} for a system of spinless point particles, using
a general construction of a complete series of time ordered Cauchy
surfaces, concludes that, also in the case of a closed universe, if
Gott pairs are produced
as envisaged in \cite{GuthII}, the universe collapses before a CTC
can form. Tipler \cite{Tipler} and Hawking \cite{Hawking} assuming the
weak energy condition proved in
3+1 dimensions that if CTC are formed in a compact region of space-time,
then one must necessarily have the creation of  singularities.
On the other hand it is very simple to recognize \cite{JDtH}
that a point like source
with angular momentum produce sufficiently near the source CTC.
In fact for a point like spinning particle the metric is given by
\begin{equation}
ds^2=(dt+4     J~d\theta)^2-\alpha^2\rho^2d\theta^2-d\rho^2
\end{equation}
where $J$ is the angular momentum of the source in units $c^3/G$  and
$\alpha=1-4GM$.
For $16 J^2-\alpha^2\rho^2>0$ i.e. for $\displaystyle{
\rho<\left |{{4 J}\over \alpha}\right |}$, we have
the CTC $\rho$=const $t$=const and $0\leq\theta<2\pi$. Usually the
appearance of such a CTC is ascribed to the unphysical nature of a
point spinning particle as the energy momentum tensor is singular at the
origin.
 A conjecture \cite{JDtH,Hawking} states that if the metric is
generated by a physical energy momentum tensor and proper boundary
conditions are imposed, no CTC should develop. By physical sources one usually
understands an energy momentum tensor that satisfies one or more among the
weak, dominant and  strong energy condition \cite{H.E.}.
Using the results of the previous section, we shall show here that for
a stationary source with circular symmetry satisfying the WEC, if the
universe is open, no CTC can appear provided there are no CTC at space
infinity.\\
The proof follows from direct manipulations of Einstein's equations in
the form (\ref{symmrot}) combined with the WEC and it extends immediately
to the case of cylindrical universes (Case 2 with
$\alpha^0_1=\beta^0_1=0$)
 and to linear universes (Case 1 with
$(\alpha_2^0)^2=(\beta_2^0)^2\neq 0$).\\
It is immediately  seen that a CTC implies the existence of a CTC with
constant $\rho$ and $t$. In fact if we call $\sigma$ the parameter of the
curve $(0\le\sigma<1)$,  $t(\sigma)$ must satisfy $t(0)=t(1)$ and at the point
$\sigma_0$ where $\displaystyle{{dt\over d\sigma}=0}$
we have that the tangent vector
$(0,d\theta,d\rho)$ is time like i.e. also $(0,d\theta,0)$ is time like,
and thus the
circle $t=0$,
$\rho=\rho(\sigma_0)$ is time-like. Thus to prove that CTC cannot exist
it is sufficient to show
that $g_{\theta\theta}$, which by assumption is negative at space
infinity, cannot change sign. \\
To begin if the determinant of the dreibeins  in the reduced radial
gauge vanishes at a certain
$\bar\rho$, it means that the manifold at $\rho=\bar\rho$ either
closes or becomes singular.
In fact let us consider the trace of the energy momentum tensor, which
is an invariant. From eq. (\ref{tabrot}) we obtain
\begin{equation}
T^\mu_\mu= -{1\over \kappa}[{(\det (e))''\over \det (e)}+
{\alpha_1\beta_2 -\alpha_2\beta_1 \over \det (e)}]
\end{equation}
where $T^\mu_\mu$ is the trace of the energy momentum tensor,  which
is related to ${\cal T}^{ab}$ by
$T_{\mu\nu}=\rho\displaystyle{\frac{{\cal T}^{ab} e_{a\mu}
e_{b\nu}}{{\rm det}(e)}}$. $T^\mu_\mu$ is an invariant
that we assume according to our general hypothesis to be regular.\\
We recall that $T^\mu_\mu=-\displaystyle{\frac{1}{2\kappa}}~R$.
We notice that the second term of the r.h.s. is the eigenvalue
$\lambda_2$ of $T^{ab}$ and as such also an invariant. As a consequence
$\displaystyle{(\det (e))''\over \det (e)}$ is
 also an invariant and a regular function
of $\rho$, according to our general assumption. Then if $\det( e)$
vanishes like
$c r^\alpha$ with $r=\bar\rho -\rho$ we have
$\alpha(\alpha-1)r^{-2}$=regular
which
implies either $\alpha=0$ (then $\det (e)$ does not vanish) or $\alpha=1$.
In more detail, solving
\begin{equation}
\label{regularity}
{(\det (e))''\over \det (e)}=f(r)={\rm regular~function~of }~~r
\end{equation}
one finds $\det(e)= c~ r (1+O(r^2))$.
We distinguish two cases\\

\noindent
{\bf 1.}
$A_2$ and/or $B_2\neq 0$  in $\bar \rho$. Then  there  always exists a
rotation which makes  $A_1=B_1=0$ in $\bar\rho$. Thus without loss of
generality we can consider the case $A_2$ and/or $B_2\ne 0$ and
$A_1=B_1=0$.  Then in a neighbourhood of $\bar \rho$  the
metric becomes
\begin{equation}
ds^2= r^2 (\alpha_1^2-\beta_1^2)  dt^2+2 r^2
(\alpha_1\alpha_2-\beta_1\beta_2) d\theta dt+(A^2_2-B^2_2)d\theta^2
-dr^2.
\end{equation}
If we impose, as required by eq. (\ref{regularity}), that ${\rm det}(g)=c
r^2(1+O(r^2))$ we must have in $\bar\rho$,\
$(\alpha^2_1-\beta^2_1)(A^2_2-B^2_2)\ne 0$
and negative as required by eq. (\ref{symmrot}). If
$\alpha_1^2-\beta^2_1>0$ and $A_2^2-B_2^2<0$ using the following
transformation
\begin{equation}
\label{xx}
\tau=r\sinh\sqrt{\alpha_1^2-\beta_1^2}~t,\ \ \ \
\zeta=r\cosh\sqrt{\alpha_1^2-\beta_1^2}~t
\end{equation}
we reduce the metric to the regular form
\begin{equation}
ds^2=d\tau^2-d\zeta^2+(A_2^2-B_2^2)d\theta^2+
2\frac{(\alpha_1\alpha_2-\beta_1\beta_2)}{\sqrt{\alpha_1^2-\beta_1^2}}
(\zeta d\tau-\tau d\zeta)d\theta.
\end{equation}
But at the events $\tau=0,\ \zeta=0,$ due to (\ref{xx})
we have only space-like
displacements and thus the manifold is singular. If on the other hand
$\alpha_1^2-\beta^2_1<0$ and $A_2^2-B_2^2>0$ using the following
transformation
\begin{equation}
x=r\cos\sqrt{\beta_1^2-\alpha_1^2}~t,\ \ \ \
y=r\sin\sqrt{\beta_1^2-\alpha_1^2}~t
\end{equation}
the metric takes the regular form
\begin{equation}
ds^2=-dx^2-dy^2+(A_2^2-B_2^2)d\theta^2+
2{(\alpha_1\alpha_2-\beta_1\beta_2)\over \sqrt{\beta_1^2-\alpha^2_1}}
(x dy-y d x)d\theta.
\end{equation}
The only possibility to have a manifold in the neighbourhood of
$x=y=0$ is that $t$ is a periodic variable with period
$\displaystyle{2\pi\over\sqrt{\beta^2_1-\alpha^2_1}}$. i.e. the slice
$\rho={\rm const}$ would assume for $r$ small and positive  the topology of
$S_1\times S_1$. But the topology near the origin $\rho=0$ is $R\times
S_1$ and as ${\rm det}(e)=g_{00} g_{\theta\theta}-g_{0\theta}^2$ does
not vanish between $0$ and $\bar\rho$ such a change  of topology is
not possible. Thus in $r=0$ the manifold is singular.

\noindent{\bf 2.} $A_2=B_2=0$ in $\bar\rho$.  The metric for small $r$
becomes
\begin{equation}
ds^2=(A_1^2-B_1^2) dt^2+ 2 r^2 (\alpha_1\alpha_2-\beta_1\beta_2) dt d\theta
+r^2 (\alpha_2^2-\beta^2_2) d\theta^2-dr^2
\end{equation}
 and for the same reason as in case 1 $(A_1^2-B_1^2)
(\alpha_2^2-\beta_2^2)<0$. Performing the rotation
$\theta=\theta^\prime-\displaystyle{\alpha_1\alpha_2-\beta_1\beta_2\over
\alpha^2_2-\beta_2^2} t$ we obtain
\begin{equation}
ds^2= ( A_1^2-B_1^2+ O(r^2)) dt^2 + r^2 (\alpha^2_2-\beta^2_2)
{d\theta^\prime}^2-dr^2.
\end{equation}
For $A_1^2-B_1^2>0$ the universe closes like a cone in
$\rho=\bar\rho$. Such a closure may be regular if
$\alpha_2^2-\beta_2^2=1$.
 For $A_1^2-B_1^2<0$ by means of the transformation
\begin{equation}
x=r \cosh\sqrt{\alpha_2^2-\beta^2_2}\ \theta\ \ \ \ \
y=r \sinh\sqrt{\alpha_2^2-\beta^2_2}\ \theta
 \end{equation}
identifying the points with $\theta=\pi$ with those with
$\theta=-\pi$, we have the regular metric
\begin{equation}
ds^2=c^2 dt^2 -dx^2-dy^2.
\end{equation}
But as
$-\pi\le\theta\le\pi$, we have in $x=y=0$ only space-like
displacements and thus the manifold is singular.

The conclusion is that if ${\rm det}(e)(\bar\rho)=0$ either the
geometry is singular (divergence of the eigenvalues of the energy
momentum tensor), or the manifold becomes singular, or the universe
closes at $\rho=\bar\rho$.

\noindent
We now show that the WEC combined with $\det(e)>0$ and the absence of
CTC at infinity, implies the absence of CTC for any $\rho$. In fact the
validity of the WEC $v_a\tau^{ab} v_b\geq 0$ for the vectors
$T^a+\Theta^a$ and $T^a-\Theta^a$ gives from eq.(\ref{tabrot})
\begin{equation}
\frac{d}{d\rho} \left [ (\alpha_2\pm\beta_2)(B_1\pm A_1)-
                 (B_2\pm A_2)(\alpha_1\pm\beta_1)\right ]\ge 0
\end{equation}
that can be integrated  to give
\begin{eqnarray}
\label{WECdelta}
E^{(\pm)} (\rho)\equiv
(B_2\pm A_2)(\alpha_1\pm\beta_1)-
&&(\alpha_2\pm\beta_2)(B_1\pm A_1)\ge\nonumber\\
&&(B^0_2\pm A^0_2)(\alpha^0_1\pm\beta^0_1)-
(\alpha^0_2\pm\beta^0_2)(B^0_1\pm A^0_1)
\end{eqnarray}
where $E^\pm (\rho )$ does not depend on $\rho$ for $\rho>\rho_0$.
We study now the sign of $E^\pm(\rho_0)$.

\noindent
\noindent i) \ $(\alpha^0_2)^2-(\beta^0_2)^2\not = 0$  (conical universe).\\
Under the hypothesis that there are no CTC at infinity we have
\begin{equation}
\label{B}
(\alpha^0_2)^2-(\beta^0_2)^2<0,
\end{equation}
while the support equation for $\tau^{\rho\rho}$ is
\begin{equation}
(\alpha^0_1+\beta^0_1)(\alpha^0_2-\beta_2^0)-
(\alpha^0_1-\beta^0_1)(\alpha^0_2+\beta_2^0)=0.
\end{equation}
Using the symmetry equation (\ref{symmrot}) written in the form
\begin{eqnarray}
\label{E}
(A_2+B_2)(\alpha_1-\beta_1)+&&(A_2-B_2)(\alpha_1+\beta_1)
-\nonumber\\
&&(A_1+B_1)(\alpha_2-\beta_2)-(A_1-B_1)(\alpha_2+\beta_2)=0
\end{eqnarray}
and the determinant written as
\begin{equation}
{\rm det}(e)=\frac{1}{2} [ (A_2-B_2)(A_1+B_1)-(A_1-B_1)(A_2+B_2)],
\end{equation}
the r.h.s. of (\ref{WECdelta}) becomes
\begin{equation}
E^{(\pm)}(\rho_0)=-\frac{\alpha^2_0\pm\beta^0_2}{\alpha^0_2\mp\beta_2^0}
\frac{d}{d\rho} {\rm det}(e)\mid_{\rho=\rho_0}
\end{equation}
that due to (\ref{B}) and
$\displaystyle{\frac{d {\rm det}(e)}{d\rho}\mid_{\rho=\rho_0}\ge 0}$
is non negative i.e. $E^{(\pm)}(\rho_0)\ge 0$.\\
\medskip
\noindent ii) $\alpha^0_2=\beta^0_2\not=0$ (linear universe).\\
Using the support equation we have $\alpha_1^0=\beta_1^0$ which implies
$E^{(-)}(\rho_0)=0$. From the equation of motion (\ref{E}) and
${\rm det}(e)\not =0$ one obtains $A^0_2-B^0_2\not =0$ and thus we have
\begin{equation}
E^{(+)}(\rho_0)=-2\frac{\alpha_2^0+\beta^0_2}{A^0_2-B^0_2} {\rm det }(e).
\end{equation}
But ${\rm det}(e)>0$ while the sign of $\displaystyle{
\frac{\alpha_2^0+\beta^0_2}{A^0_2-B^0_2}}
$ has to be negative if there are no CTC at infinity, because $A^2_2-B^2_2$
in this case behaves linearly at infinity.\\
Similarly for $\alpha^0_2=-\beta^0_2$ we obtain $E^{(+)}(\rho_0)=0$
and $E^{(-)}(\rho_0)>0$.\\
\medskip
\noindent iii) If $\alpha^0_1=\alpha_2^0=\beta^0_1=\beta^0_2=0$
(cylindrical universe)
we have also $E^{(\pm)}(\rho_0)=0$.
\\
The only case that escapes our analysis is
$\alpha_2^0=\beta^0_2=0$ and $\alpha^0_1$ and/or $\beta^0_1$ $\not = 0$;
this situation corresponds to a cylindrical universe generated by a string
with tension and total angular momentum 0. Thus except for this case  we
have that the r.h.s. of equation (\ref{WECdelta}) is
$E^{(\pm)}(\rho_0)\ge 0$.\\
Let us now consider the following combination
\begin{equation}
(A_2-B_2)^2 E^{(+)}(\rho)+(A_2+B_2)^2 E^{(-)}(\rho)\ge 0.
\end{equation}
A little algebra shows that the l.h.s. equals
\begin{equation}
-2 {\rm det}(e)^2 \frac{d}{d\rho} \left (\frac{ A^2_2(\rho)-
B^2_2(\rho)}{{\rm det}(e)}\right ).
\end{equation}
Thus we reached the conclusion that $\displaystyle{\frac{d}{d\rho}
\left (\frac{ A^2_2(\rho)-
B^2_2(\rho)}{{\rm det}(e)}\right )}\le 0$; it means  that
$\displaystyle{
\left (\frac{ A^2_2(\rho)-
B^2_2(\rho)}{{\rm det}(e)}\right )}$ is a non increasing function
of $\rho$. As $\displaystyle{
\left (\frac{ A^2_2(\rho)-
B^2_2(\rho)}{{\rm det}(e)}\right )}$  at the origin is zero we obtain
that $A^2_2(\rho)-
B^2_2(\rho)$ is always negative and thus we cannot have CTC.\\
In the case of a regular open universe
the hypothesis that there are no CTC at infinity and that at least an
average form of the WEC is satisfied (see eq.(\ref{WECdelta})),  are not only
sufficient but also necessary for the absence of CTC.
 In fact we give now examples of
regular sources that violate the WEC and produce  CTC for finite radius  but
no CTC at infinity. In addition we shall construct examples  of  sources
 satisfying the WEC and produce CTC for any  radius  larger than a certain
$\rho_1$. \\
With regard to the first example we consider the following functions
\begin{equation}
\alpha_1=0,\  \ \ \ \ \ \ \ \ \ \ \ \ \beta_2=1
\end{equation}
and due to the eq. (\ref{symmrot})
\begin{equation}
\label{ccc}
\alpha_2=-B_1+\rho \beta_1.
\end{equation}
The metric becomes
\begin{equation}
ds^2=(1-B^2_1)d t^2-2 (A_2+\rho B_1)d t d \theta+ (A^2_2-\rho^2)d\theta^2-
d\rho^2.
\end{equation}
In addition we choose $\beta_1(0)=0$, $\beta_1(\rho)=0$ for $\rho>\rho_0$ and
$\displaystyle{\int^{\rho_0}_0
\beta_1(\rho) d\rho}=0$ i.e. $B_1(\rho)=0$ for $\rho>\rho_0$. We have
\begin{eqnarray}
&&\int^{\rho_0}_0 \alpha_2(\rho)\ d\rho=A_2(\rho_0)=
\int^{\rho_0}_0 (-B_1+\rho \beta_1)\ d\rho=
\rho_0 B_1(\rho_0)-2\int^{\rho_0}_0 B_1(\rho)\ d\rho=\nonumber\\
&&-2\int^{\rho_0}_0 B_1(\rho)\ d\rho
\end{eqnarray}
that we shall choose different from zero. Then  the exterior metric
($\rho>\rho_0$) is
\begin{equation}
ds^2=(d t-A_2(\rho_0) d \theta)^2-\rho^2 d\theta^2-
d\rho^2,
\end{equation}
and it obviously possesses CTC for $\rho< A_2(\rho_0)$. By properly taking
the normalization of $\beta_1$ we can always have $|A_2(\rho_0)|
>\rho_0$.  The proved theorem tell us that our source must violate the WEC.
This can be also
directly seen from the fact that ${\cal T}^{00}=-\displaystyle{
\frac{1}{\kappa\rho} A_2\beta^\prime_1}$ changes sign at the point
where $\beta^\prime_1$
reaches the first zero starting from the origin because there
$A_2(\rho)$ has a well defined sign due to the fact that $\alpha_2$ is
monotonic up to the first zero of $\beta_1^\prime$ (from (\ref{ccc}) we
have $\alpha_2^\prime=\rho
\beta_1^\prime$ and  $\beta^\prime_1$ must
possess a zero for  $\rho$ belonging to
$[0,\rho_0]$ because $\beta_1(0)=\beta_1(\rho_0)=0$).\\
We come now to the second example  i.e. of a source that satisfies the WEC
and generates a metric with CTC at infinity. To this purpose we take
$\alpha_1=0$, $\beta_1=1$, $\beta_2=1+f^\prime (\rho)$ with
$f^\prime (\rho)=o(\rho)$ and
$f^\prime (\rho)=-1$ for $\rho\ge \rho_0=1$. Then
\begin{equation}
\alpha_2=f-\rho f^\prime \ \ \ , \ \ \ A_2=2 F-\rho f\ \ \ {\rm and}
\ \ \ B_2=\rho+f
\end{equation}
with $f(\rho)=\displaystyle{\int^\rho_0 f^\prime(\rho)\  d\rho}$
and
$F(\rho)=\displaystyle{\int^\rho_0 f(\rho)\  d\rho}$.
The energy momentum tensor has the
following form
\begin{equation}
\label{ex2}
\begin{array}{ccc}
{\cal T}^{00}=-\displaystyle{\frac{1}{\kappa\rho}}
  f^{\prime\prime}(\rho) &\ \ , \ &{\cal T}^{0\theta}=\displaystyle{
\frac{1}{\kappa\rho}}
 \rho f^{\prime\prime}(\rho) \\
\ \ {\cal T}^{\theta\theta}=-\displaystyle{\frac{1}{\kappa\rho}} \ \ \ \
 \rho^2 f^{\prime\prime}(\rho) &\ \ \ \ \ \ \ \ \ \ , \ \ \ \ \ \ \ &
{\cal T}^{\rho\rho}=\displaystyle{\frac{1}{\kappa\rho}}
 (f(\rho)-\rho f^{\prime}(\rho)),\\
\end{array}
\end{equation}
and the other elements equal to zero. The eigenvalues and relative
eigenvectors are
\begin{eqnarray}
&&\lambda_0=-\frac{1}{\kappa\rho}f^{\prime\prime}(\rho)(1-\rho^2)\ \ , \ \
v^0=(1,-\rho,0);\\
&&\lambda_1=0\ \ \ \ \ \ \ \ \ \ \ \ \ ,\ \ \ \ \ \ \ \ \ \ \ \ \
 v^1=(\rho,-1,0);\\
&&\lambda_2=\frac{1}{\kappa\rho}(\rho f^{\prime}(\rho)-f(\rho))\ \ , \ \
v^2=(0,0,1).
\end{eqnarray}
For $f^{\prime\prime}(\rho)\le 0$ $\lambda_0\ge 0$ thus
$\lambda_0\ge\lambda_1$. The $\lambda_2$ is negative because
$\lambda^\prime_2=\rho f^{\prime\prime}\le 0$ and $\lambda_2(0)=0$. Thus
$\lambda_1\ge \lambda_2$ and we have satisfied the WEC \cite{H.E.}.
Outside the source i.e. $\rho\ge 1$ $g_{\theta\theta}$ becomes
$[A_2(1)+(\rho-1) \alpha_2(1)]^2-B_2^2(1)$ and thus as $\alpha_2(1)\not =0$
(we recall that $\alpha_2^\prime=-\rho f^{\prime\prime}$) $g_{\theta\theta}$
goes like $\alpha_2(1)\rho^2$ for $\rho\to \infty$, thus giving CTC at
infinity.
 As mentioned above
the only case of open universe  with no CTC at infinity
that
escapes our analysis is the rather unphysical situation of a cylindrical
universe with
zero angular momentum of the type generated by a closed string with tension
\cite{JDtH}
for which on the other hand no CTC exists outside the source.
\section{Conclusion}
In this paper  we developed a quadrature procedure for solving the general
stationary problem in 2+1 dimensional gravity in presence of extended
sources also endowed  with angular momentum. Such an approach reveals itself
particularly useful in treating the problem of CTC in the case of
rotational symmetry. The result is that an average form of the WEC
(which is  a consequence of the WEC itself) and the absence of
CTC at space infinity exclude the existence of CTC
everywhere for open universe, conical at infinity.
Explicit counterexamples show that such conditions are not only
sufficient but also necessary in the sense it is possible to find
sources which do not satisfy the WEC, give rise to CTC for finite
$\rho$ but generate no CTC at infinity, but also sources exist that
satisfy the WEC and give rise to CTC at infinity.
We mention that the radial gauge approach
works also for the time dependent situation \cite{MS2} and thus it appears
a good  candidate for
treating the problem of evolution of continuous distribution of
matter in 2+1 dimensional gravity.

\appendix
\section{}
It was shown in section 4 of \cite{MS1} that the gauge
\begin{equation}
\label{form1}
\sum_i\xi^i\Gamma^a_{bi}=0
\end{equation}
\begin{equation}
\label{form2}
\sum_i\xi^i e^a_i=\sum_i\delta^a_i\xi^i
\end{equation}
is attainable. We prove now formulae (3-6) of Sec.2 of the body of the paper
which play a major role in the subsequent developments.

{}From  (\ref{form1}) and (\ref{form2})
 and the regularity of $\Gamma^a_{b\mu}$ and $e^a_\mu$ at the
origin we obtain
\begin{equation}
\Gamma^a_{bi}({\bf 0})=0 ~~~~{\rm and}~~~~e^a_i({\bf 0})=\delta^a_i.
\end{equation}
{}From the definition of the components $R^a_{b\mu\nu}$ of the Riemann two
form we obtain
\begin{equation}
\xi^i R^a_{b i\mu}=\xi^i\partial_i\Gamma^a_{b\mu}+\Gamma^a_{bi}\delta^i_\mu
\end{equation}
and thus
\begin{equation}
\label{A1}
\G^a_{bi}(\bfxi)=\xi^j\int^1_0 R^a_{bji}(\lambda \bfxi)\lambda d\lambda.
\end{equation}
For $\G^a_{b0}(\bfxi)$ we have
\begin{equation}
\xi^i R^a_{bi0}=\xi^i\dev_i\G^a_{b0}
\end{equation}
and thus
\begin{equation}
\Gamma^a_{b0}({\bfxi})=\G^a_{b0}({\bf 0})+\xi^i\int^1_0 R^a_{bi0}(\l {\bfxi})
d\l
\end{equation}
where $\G^a_{b0}({\bf 0})$ is an arbitrary constant of integration. For the
vierbeins we have
\begin{equation}
\xi^i\dev_i e^a_0-\xi^i \Gamma^a_{i0}=\xi^i S^a_{i0}
\end{equation}
being $S^a_{i0}$ the torsion, from which
\begin{equation}
\label{A9}
e^a_0({\bf \xi})=\delta^a_0+\xi^i\int^1_0\Gamma^a_{i0}(\l{\bfxi}) d\l+
\xi^j\int^1_0 S^a_{j0}(\lambda\bfxi)d\l.
\end{equation}
Substituting (\ref{A1}) into (\ref{A9}) we obtain eq. (\ref{ea0}) of sect.II.
Similarly for $e^a_i$ we obtain from
\begin{equation}
\xi^i\dev_i e^a_j-\xi^i\dev_j e^a_i=\xi^i\G^a_{ij}
+\xi^i S^a_{ij}
\end{equation}
\begin{equation}
e^a_i=\delta^a_i+\xi^j\int^1_0\G^a_{ji}(\l {\bfxi})\l d\l
+\xi^j\int^1_0 S^a_{ji}(\lambda\bfxi) \l d\l
\end{equation}
from which we derive eq. (\ref{eai}) of sect.II. It is worth mentioning
that in all these
equations as in eq. (\ref{eai}) and (\ref{ea0}) of sect.II $R^a_b$
 and $S^a$ are
the Riemann and torsion two forms in the radial gauge and not in an
arbitrary gauge and thus they cannot be chosen arbitrarily but they are
subject to a reduced form of Bianchi identities analogous to those
found in ref.
\cite{MT}.
\section{}

We derive here the general conditions imposed upon the functions
$\alpha_i$, $\beta_i$ and $\gamma_i$ by the regularity at the origin
of the energy momentum tensor $\tau^\rho_c$. The simplest way to proceed is to
express the unit vectors $\Theta$ and $P$ in terms of the carthesian
versors $X$ and $Y$. Denoting with $T^{xx},T^{xy}$ etc. the carthesian
components we obtain the following behaviors
\begin{eqnarray}
\alpha_1= &&\Gamma^{01}_0({\bf 0})\cos \theta+
\Gamma^{02}_0({\bf 0})\sin \theta -\kappa
[T^{xx}\sin^2\theta+T^{yy}\cos^2\theta-\nonumber\\
&&(T^{xy}+T^{yx})\sin\theta\cos\theta]
\rho+o(\rho)
\end{eqnarray}
\begin{equation}
\beta_1=\Gamma^{12}_0({\bf 0})-\kappa
(-T^{0x}\sin\theta+T^{0y}\cos\theta)\rho+o(\rho)
\end{equation}
\begin{eqnarray}
\gamma_1= &&\Gamma^{01}_0({\bf 0})\sin \theta-
\Gamma^{02}_0({\bf 0})\cos \theta -\kappa
[T^{xy}\cos^2\theta-T^{yx}\sin^2\theta+\nonumber\\
&&(T^{yy}-T^{xx})\sin\theta\cos\theta]
\rho+o(\rho)
\end{eqnarray}
\begin{equation}
\alpha_2=-{\kappa\over 2}(-T^{x0}\sin\theta+T^{y0}\cos\theta)\rho+o(\rho^2)
\end{equation}
\begin{equation}
\beta_2=1-{\kappa\over 2}T^{00}\rho^2+o(\rho^2)
\end{equation}
\begin{equation}
\gamma_2=-{\kappa\over 2}(T^{x0}\cos\theta+T^{y0}\sin\theta)\rho^2+o(\rho^2)
\end{equation}

\section{}

Eqs. (\ref{ridcompattezza}) and (\ref{CC}) of the text describe the
compactness of the support
 of the energy momentum tensor $\tau^\rho_c$. We show here that
necessary and sufficient
condition for them to be satisfied is
\begin{equation}
\label{B1}
\alpha_1 B_1-\beta_1 A_1= {\rm constant}~~~~{\rm for}~~~~ \rho>\rho_0(\theta)
\end{equation}
and
\begin{equation}
\label{B2}
\alpha_1^2-\beta_1^2+\gamma_1^2=
{\rm constant}~~~~{\rm for}~~~~ \rho>\rho_0(\theta)
\end{equation}
where $\alpha_1,\beta_1,\gamma_1$ become all independent of $\rho$ for
$\rho>\rho_0(\theta)$. The constants appearing in (\ref{B1}) and (\ref{B2})
 do not
depend both on $\rho$ and  $\theta$.

First we prove  (\ref{B1}) and (\ref{B2}) starting  from
(\ref{ridcompattezza}) and (\ref{CC}).
In fact multiplying (\ref{ridcomp1}) by $-\beta_1$, (\ref{ridcomp2})
by $\alpha_1$ and (\ref{ridcomp3}) by
$-\gamma_1$ and then summing we obtain
\begin{equation}
{\partial\over\partial\theta}(\alpha_1^2-\beta_1^2+\gamma_1^2)=0~~~~{\rm
for}~~~~\rho>\rho_0(\theta).
\end{equation}
But as for $\rho>\rho_0(\theta)$, $\alpha_1,\beta_1,\gamma_1$ do not
depend on $\rho$  we have equation (\ref{B2}).

We come now to the proof of eq. (\ref{B1}). For generic functions
$\alpha_i,\beta_i,\gamma_i$ that for $\rho>\rho_0(\theta)$ do not
depend on $\rho$, the necessary
and sufficient condition for having $\alpha_2^\prime=0$  for
$\rho>\rho_0(\theta)$, is the constancy in $\rho$ of
$A_2\alpha_1-A_1\alpha_2$. In fact for $\rho>\rho_0(\theta)$ $A_i$
become
$A_i=\alpha^0_i(\rho-\rho_0)+A_i^0$.
{}From eq. (\ref{Eq27}) this means that
\begin{equation}
\label{B3}
A_1^2{\partial\over\partial\rho}\left ({A_2\over A_1}\right )=
{A_1^2B_1^2\over B_1^2-A_1^2} {\partial\over\partial\rho}({N\over
A_1B_1})
\equiv K(\rho,\theta)
\end{equation}
does not depend on $\rho$ for $\rho>\rho_0(\theta)$. Performing
explicitly the derivative on the r.h.s. and substituting $N$ given by
(\ref{N}) into (\ref{B3}) and taking into account that $\gamma_1$ does not
depend on $\rho$ for $\rho>\rho_0(\theta)$, we obtain always for
$\rho>\rho_0(\theta)$
\begin{equation}
\label{C5}
A_1^2(B_1{\partial\alpha_1\over\partial\theta}-\beta_1{\partial
A_1\over\partial\theta})-
B_1^2(A_1{\partial\beta_1\over\partial\theta}-\alpha_1{\partial
B_1\over\partial\theta})= {H(\theta)\over 2} (B_1^2-A_1^2).
\end{equation}
For $\rho>\rho_0(\theta) $ both members of (\ref{C5}) become second
degree polynomials in $\rho$. Equating the coefficients  we reach,
always for $\rho>\rho_0(\theta)$,
\begin{equation}
B_1{\partial\alpha_1\over\partial\theta}-
\beta_1{\partial A_1\over\partial\theta}=
A_1{\partial\beta_1\over\partial\theta}-
\alpha_1{\partial B_1\over\partial\theta}\equiv-\frac{ H(\theta)}{2}
\end{equation}
i.e.
\begin{equation}
\label{B7}
{\partial\over\partial\theta}(B_1\alpha_1-A_1\beta_1)=0.
\end{equation}
Viceversa
$\alpha_1^{'}=\beta_1^{'}=\gamma_1^{'}=0$ and (\ref{B1}) and
(\ref{B2}) imply the compactness conditions. In fact from eq.
(\ref{B7}) going back we have that $A_2\alpha_1-A_1
\alpha_2=f(\theta)$ i.e. $\alpha^{'}_2=0$ for $\rho>\rho_0(\theta)$.
{}From eq. (\ref{22a}) we have now $\beta^\prime_2=0$. Multiplying (\ref{22b})
by $\beta_1$ and (\ref{22c}) by $\alpha_1$ and subtracting we obtain
\begin{equation}
\label{CC6}
\gamma_2^\prime (\beta_1 A_1-\alpha_1 B_1)=
\gamma_1(\alpha_2\beta_1-\alpha_1\beta_2)+
\frac{1}{2}\frac{\partial}{\partial\theta}
( \beta_1^2-\alpha_1^2).
\end{equation}
But due to eq. (\ref{N}), always for $\rho>\rho_0 (\theta)$, the
r.h.s. of (\ref{CC6}) vanishes and thus $\gamma^\prime_2=0$. The vanishing
of $\gamma_i^\prime$  via eqs. (\ref{22b}) and (\ref{22c}) implies eqs.
(\ref{ridcomp1}) and (\ref{ridcomp2}).  From (\ref{B2}) and
the vanishing of the r.h.s.
of (\ref{CC6})  we obtain
\begin{equation}
\gamma_1
(\alpha_1\beta_2-\alpha_2\beta_1)-\gamma_1\frac{\partial\gamma_1}{\partial
\theta} =0
\end{equation}
i.e. (\ref{ridcomp3}).

\end{document}